\documentclass{article}

\usepackage{microtype}
\usepackage{graphicx}
\usepackage{subfigure}
\usepackage{booktabs} 

\usepackage{hyperref}

\usepackage[accepted]{icml2023}

\usepackage{amsmath}
\usepackage{amssymb}
\usepackage{mathtools}
\usepackage{amsthm}

\usepackage[capitalize,noabbrev]{cleveref}

\theoremstyle{plain}

\theoremstyle{definition}

\theoremstyle{remark}

\usepackage[textsize=tiny]{todonotes}

\icmltitlerunning{Quadtree features for machine learning on CMDs}

\begin{document}

\twocolumn[
\icmltitle{Quadtree features for machine learning on CMDs}



\icmlsetsymbol{equal}{*}

\begin{icmlauthorlist}
\icmlauthor{J. Schiappacasse-Ulloa}{equal,sch}
\icmlauthor{M. Pasquato}{equal,sch,yyy,melone,pera}
\icmlauthor{S. Lucatello}{comp}
\end{icmlauthorlist}

\icmlaffiliation{sch}{Dipartimento di Fisica e Astronomia, Universita’ di Padova, Vicolo dell’Osservatorio 3, I-35122, Padova, Italy.}
\icmlaffiliation{yyy}{Département de Physique, Université de Montréal, Montreal, Quebec H3T 1J4, Canada.}
\icmlaffiliation{melone}{Mila - Quebec Artificial Intelligence Institute, Montreal, Quebec, Canada}
\icmlaffiliation{pera}{Ciela, Computation and Astrophysical Data Analysis Institute, Montreal, Quebec, Canada}
\icmlaffiliation{comp}{INAF–Osservatorio Astronomico di Padova, Vicolo dell’Osservatorio 5, 35122 Padova, Italy}

\icmlcorrespondingauthor{Jose Schiappcasse-Ulloa}{joseluis.schiappacasseulloa@studenti.unipd.it}

\icmlkeywords{Machine Learning, Star Clusters, Photometry}

\vskip 0.3in
]

 

\printAffiliationsAndNotice{\icmlEqualContribution} 

\begin{abstract}
The upcoming facilities like the Vera C. Rubin Observatory will provide extremely deep photometry of thousands of star clusters to the edge of the Galaxy and beyond, which will require adequate tools for automatic analysis, capable of performing tasks such as the characterization of a star cluster through the analysis of color-magnitude diagrams (CMDs). 
The latter are essentially point clouds in N-dimensional space, with the number of dimensions corresponding to the photometric bands employed. In this context, machine learning techniques suitable for tabular data are not immediately applicable to CMDs because the number of stars included in a given CMD is variable, and equivariance for permutations is required. To address this issue without introducing ad-hoc manipulations that would require human oversight, here we present a new CMD featurization procedure that summarizes a CMD by means of a quadtree-like structure through iterative partitions of the color-magnitude plane, extracting a fixed number of meaningful features of the relevant subregion from any given CMD. 
The present approach is robust to photometric noise and contamination and it shows that a simple linear regression on our features predicts distance modulus (metallicity) with a scatter of $0.33$ dex ($0.16$ dex) in cross-validation. 
\end{abstract}

\section{Introduction}
\label{submission}

The photometric study of stellar populations typically relies on point-spread-function fitting to measure magnitudes in relevant bands. These are then combined to derive color-magnitude diagrams (CMDs) which are used to reconstruct stellar population characteristics. By comparing theoretical stellar evolutionary tracks and isochrones with
the location of CMD landmarks (in terms of color and magnitude) such as the Red Clump, Horizontal Branch, turn-off, etc. we can estimate the age, metallicity, reddening, and distance in resolved Open Clusters (OCs) and Globular Clusters \citep[GCs; e.g. ][]{2013osp..book.....C}. Properties, such as the width of the evolutionary sequences or turn-off broadening, measure binary fraction or dispersion in age, metallicity, and rotational velocity \citep[e.g.][]{2012A&A...540A..16M}. 
 
Currently, the highest quality CMDs are derived from Hubble Space Telescope (HST) photometry which, however, was far limited to very small sections of the clusters given the small field of view (FOV; ~$2$x$2$ arcmin) and WFC3 --HST's most advanced camera-- that has a magnitude limit of up to V$\sim$25.5. On the other hand, the upcoming facility Vera C. Rubin Observatory will have a FOV of $9.6$ square degrees and a magnitude limit of $27.5$ in the r band over most of the southern hemisphere. Then, it is expected to yield accurate turn-off photometry of all star clusters in its survey volume out to the edge of the Milky Way (MW). Alongside opportunities, the volume of data (roughly $20$~TB/night) collected will bring extraordinary challenges in data handling, and developing new approaches to reduction and analysis strategies.

In particular, the most common approaches to studying CMDs of stellar clusters are optimized for much smaller datasets and hence samples of stars which generally require a high degree of human intervention in parsing and examining the data. These present substantial drawbacks: high latency on large datasets, some subjectivity in the CMD landmark definitions, and in turn on properties measurement, in the outlier treatment, etc. The latter issue has been recently addressed by several authors by using a Bayesian approach \citep[][]{2011MNRAS.411..435B}, however, it does not fully remove the need to define the location of CMD landmarks and it is time-demanding. Other techniques have been proved, such as the \texttt{ASteCa} package \citep{2015ascl.soft05002P}, nevertheless, it is not time-efficient and it presents problems handling CMDs with differential reddening.

In this context, we present a new method for calculating a set of numeric features from a given CMD with the main goals of speed and robustness to deviations from ideal conditions. Our approach is based on recursively partitioning the CMD plane splitting the relevant coordinate -color or magnitude- at its median. This bears a close similarity with the computation of a quadtree data structure. It can be easily extended to higher dimensions -for instance in the case of multi-band photometry. 
\subsection{State of the art}
CMDs are the result of extracting photometric information from images. The latter can in principle be directly fed into a convolutional neural network \citep[see e.g.][]{2021MNRAS.504.5656C}, but CMDs have superior interpretability for astronomers and fit well into established data-analysis pipelines. Arguably, the first widely adopted architecture able to deal with point cloud data with the characteristics of the typical CMD is the deep set architecture \citep[][]{2017arXiv170306114Z}. Looking for citations to this paper within the Astronomy Library at the Harvard abstract service returns a grand total of nine papers (on Wed, Jun 14, 2023). All of these are in cosmology, except for \citet{2023arXiv230308474V}, which applies to membership determination in OCs using Gaia DR3 data \citep{2022arXiv220800211G}. In general, deep sets and other deep learning approaches, such as graph neural networks \citep[see][for a review]{9046288} are not built from the ground up for interpretability. It is thus understandable that astronomers look for interpretable featurization of CMDs; nonetheless, to be useful, these must be computed efficiently and without ad-hoc human interventions. 

\begin{figure*}
    \centering
    \begin{tabular}{cc}
    \includegraphics[width=0.49\textwidth]{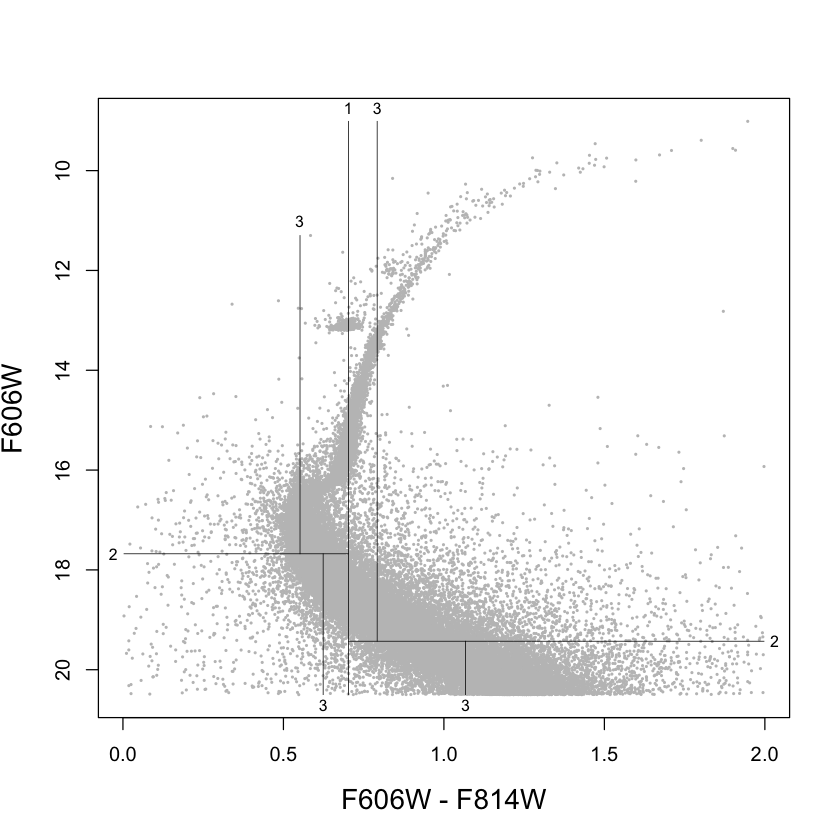} &
    \includegraphics[width=0.49\textwidth]{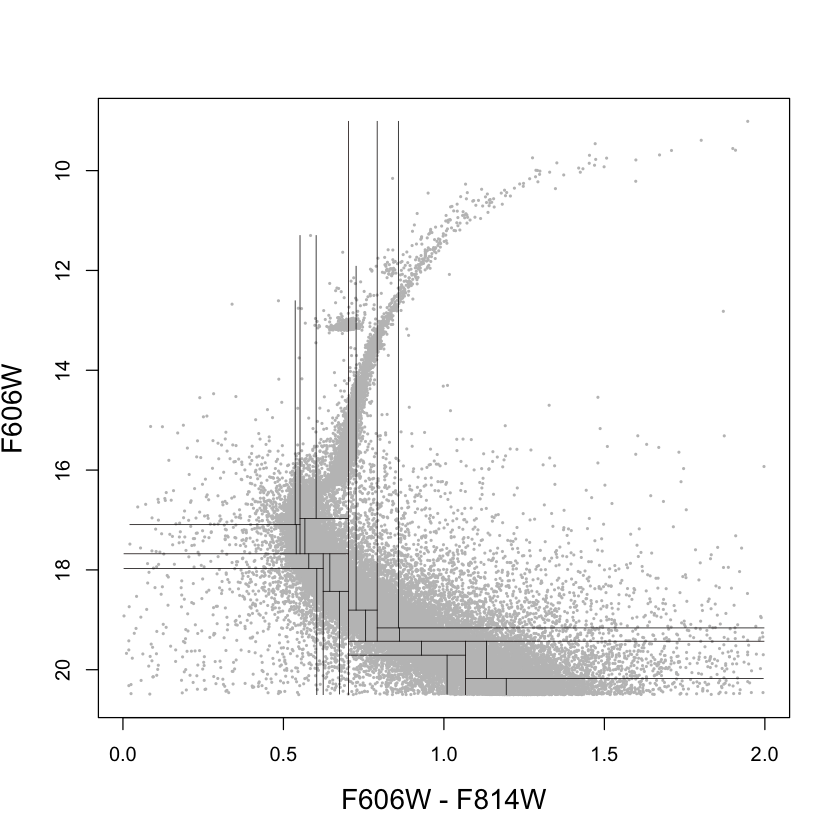}\\ 
    \includegraphics[width=0.49\textwidth]{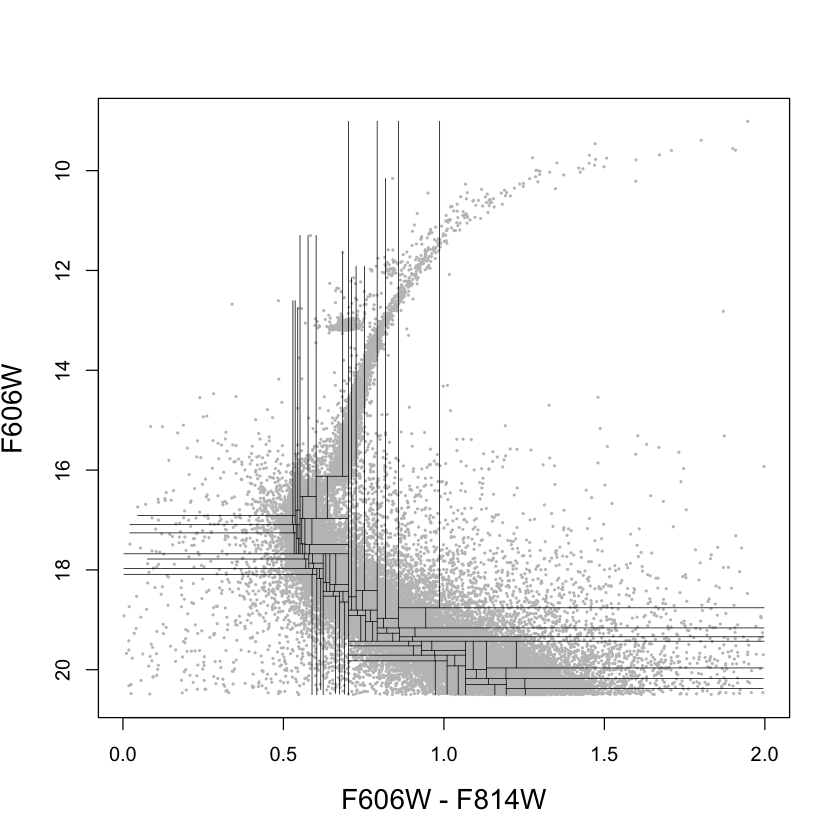} & 
    \includegraphics[width=0.49\textwidth]{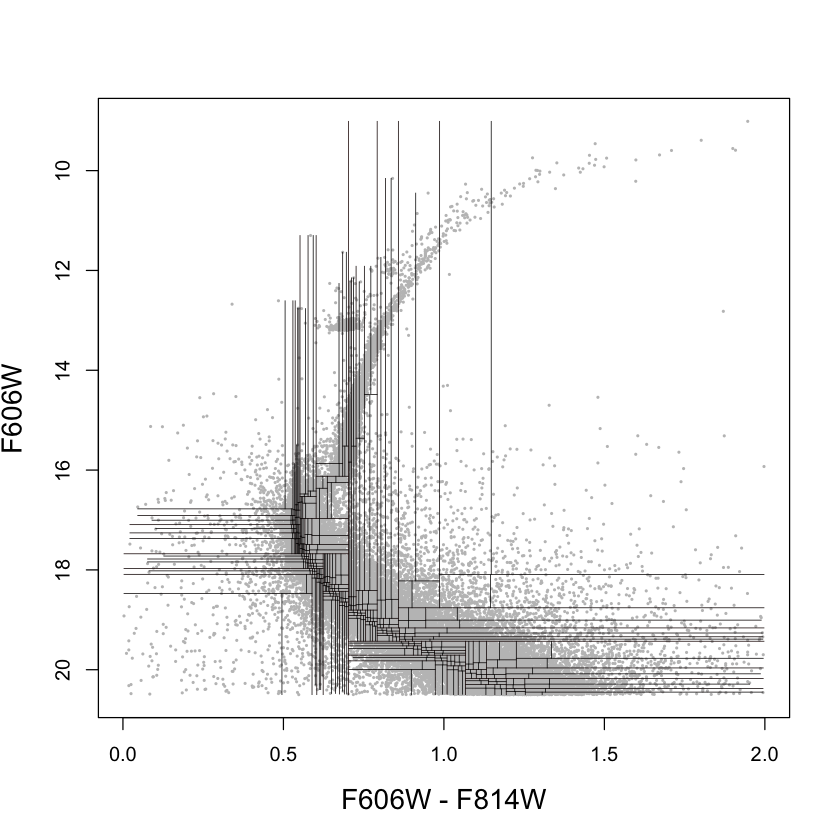}\\
    \end{tabular}
    \caption{Graphical illustration of our feature extraction procedure on the CMD of NGC 104. From the top left to bottom right the depth increases from $3$ to $9$. The lines represent the medians and are extended to the range of the data in each leaf of the quadtree. They are labeled with the respective depth in the top left panel: for instance, the vertical line labeled 1 corresponds to the median color of the CMD, the two horizontal lines labeled 2 correspond to the median magnitudes of the stars to the left and to the right of the line labeled 1, etc.} 
    \label{illustration}
\end{figure*}

\section{Data}
\label{data}
As a benchmark for our featurization approach, we selected the photometric sample from the HST Treasury GO10775 “An ACS Survey of Galactic Globular Cluster” \citep[][]{sarajedini07}, in its release through the database from the HST Treasury Program GO 13297 \citep[][]{piotto2015}. For details of the observations and data reduction, we refer the interested reader to \citet[][]{nardiello2018}. We retrieved photometric catalogs in the ACS F606W and F814W bands for a total of 56 GCs, with [Fe/H] ranging from $\sim-2.3$ to $\sim-0.3$\,dex, distance modulus (DM) (m-M)$_V$ from 12.3 to 17.6\,mag and reddening E(B-V) from negligible to $\sim0.7$\,mag. To minimize the field contamination we considered as a cluster member, stars with a membership probability higher than $0.95$ \citep[based on][]{nardiello2018}.

\begin{figure}
    \centering
    \begin{tabular}{c}
    \includegraphics[width=\columnwidth]{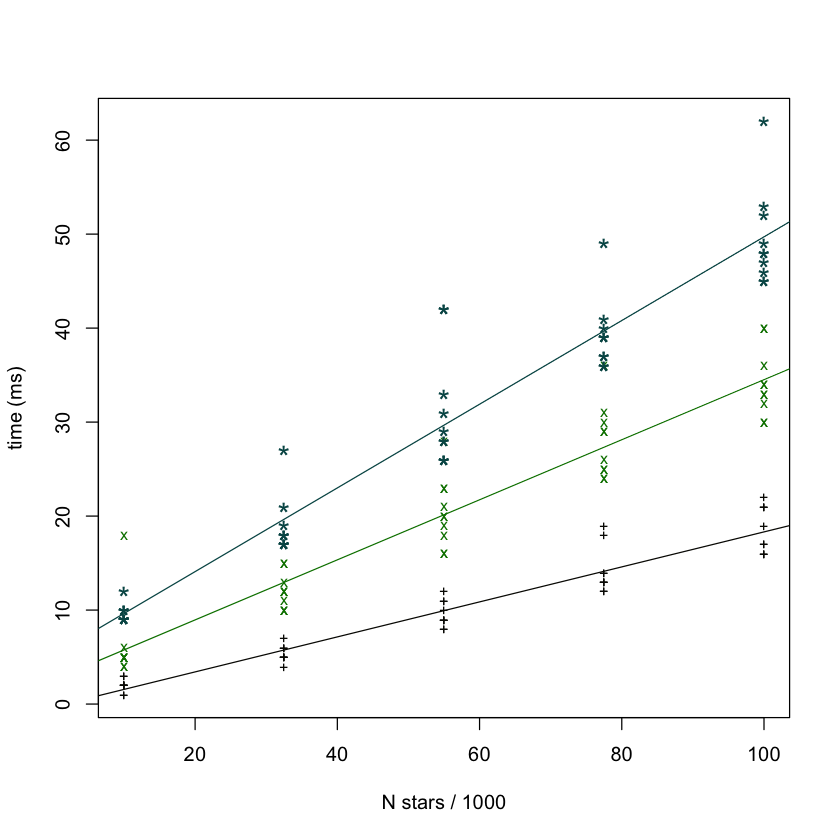}
    \end{tabular}
    \caption{Wall time required to compute our features as a function of the number of stars in the CMD (in units of 1000 stars), with an R implementation as a recursive function on an early 2015 Macbook Air. Black, green, and blue symbols correspond to runs with a depth of 3, 5, and 7 respectively. The superimposed solid lines represent a linear regression to each group.} %
    \label{scalingtime}
\end{figure}

\section{Method}
To convert CMDs into a fixed set of features, 
we calculate the median color and then split the CMD into two parts based on this median. Our featurization approach is rendered in pseudocode in algorithm \ref{alg:algo1} and illustrated visually in Fig.~\ref{illustration}. For each resulting half, we repeat the process and calculate the median magnitude. This process is repeated recursively --by the function \texttt{f}-- until a desired level (fixed by the user) of detail is reached. 
Because each split divides the number of points by two, increasing the depth may yield features that depend on the colors and magnitudes of just a few stars. Thus the depth of the recursion, \texttt{k}, should be set with the total number of stars in the CMD in mind, which may range from $10^2$ to $10^5$ for OCs and GCs, respectively. It is worth noticing that the recursion depth defines the dimensionality of the feature vector which in the present analysis ranges from $2$ to $7$. Then, the features have a fixed size avoiding the variable size of point clouds. 

\begin{algorithm}[tb]
   \caption{Median split}
   \label{alg:algo1}
\begin{algorithmic}
   \STATE {\bfseries Input:} $x$,$y$,$k$
   \STATE {\bfseries Output:} Median Value
\end{algorithmic}
\begin{algorithmic}[1]
   \STATE $m \leftarrow$ median($x$);
   \STATE {\bfseries if} $k >$ 0 {\bfseries then}
   \STATE \hskip0.5em $m1 \leftarrow$ f($y[x < m]$, $x[x < m]$, $k - 1$)\;
   \STATE \hskip0.5em $m2 \leftarrow$ f($y[x \geq m]$, $x[x \geq m]$, $k - 1$)\;
   \STATE \hskip0.5em $returnedVals \leftarrow$ [$m$, $m1$, $m2$];
   \STATE {\bfseries return} $returnedVals$
   \STATE {\bfseries else} 
   \STATE {\bfseries return} $m$\,
\end{algorithmic}
\end{algorithm}

The median calculation drives the time complexity of the algorithm once the depth of the recursion is set. The median can be found by sorting in $O(N \log N)$ or using specialized algorithms such as \emph{quickselect} or \emph{median of medians} which reach $O(N)$ in the typical case but with a worst case of $O(N^2)$ \citep[][]{blum1973time}. The sampling error of the median decreases as $1/\sqrt{N}$, similarly to that of the mean, under the assumption of normality, even though with a slightly larger constant of $\approx 5/4$. However, a great advantage of computing medians is that they are robust to outliers, having a breaking point of $0.5$. Outliers due to contamination, blending, and photometric errors are commonplace in CMDs, making our featurization procedure particularly suitable for this application. In the following, we run a series of empirical tests to corroborate this theoretical expectation.


\begin{figure*}
    \centering
    \begin{tabular}{cc}
    \includegraphics[width=0.49\textwidth]{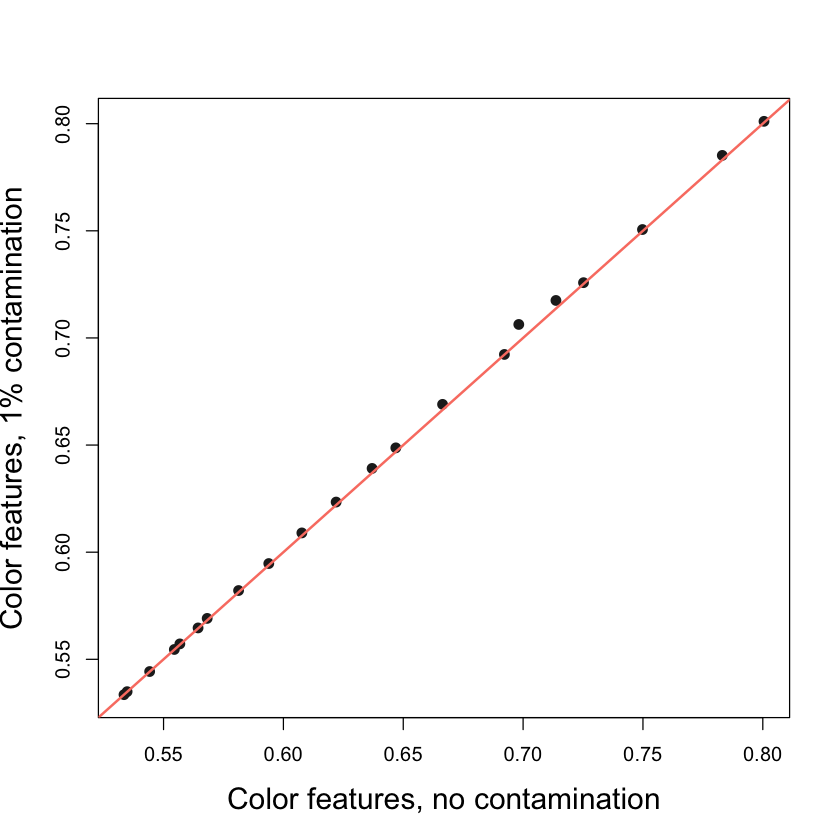} &
    \includegraphics[width=0.49\textwidth]{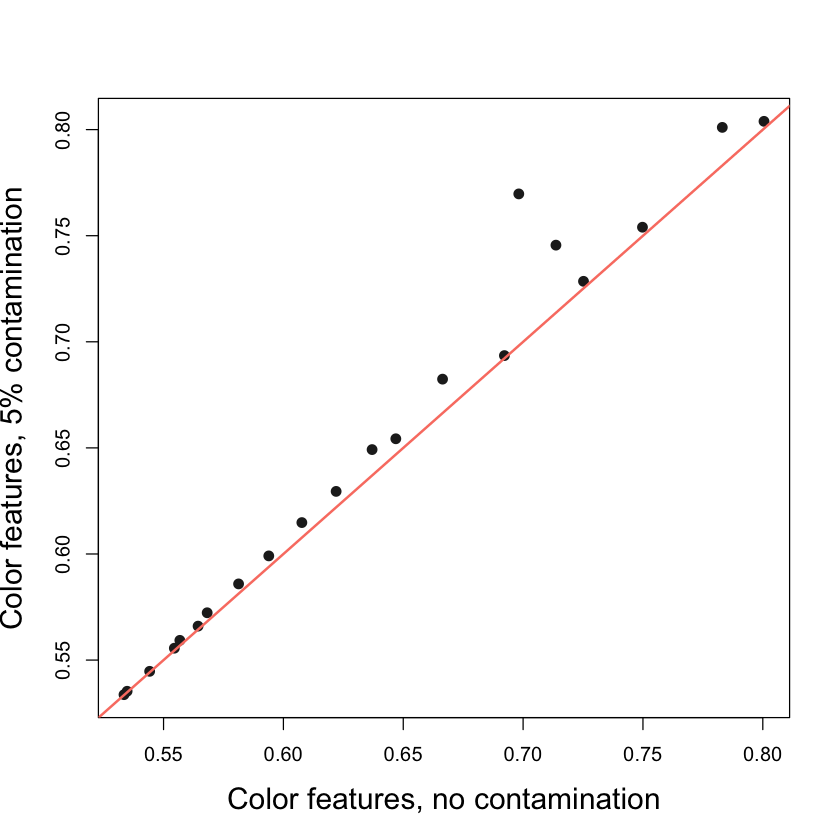} \\
    \includegraphics[width=0.49\textwidth]{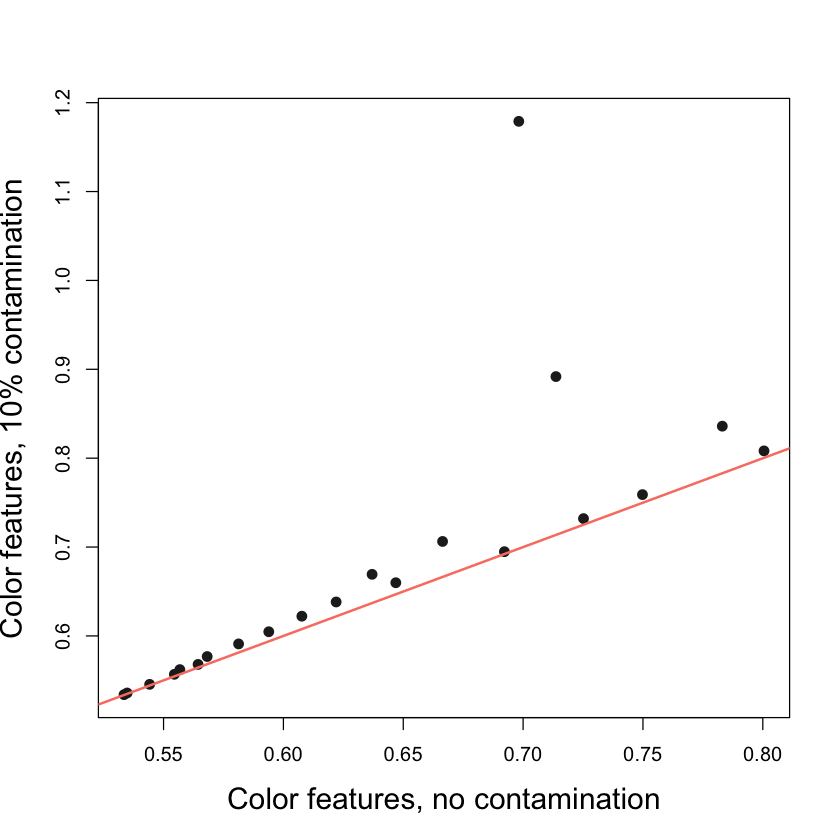}& 
    \includegraphics[width=0.49\textwidth]{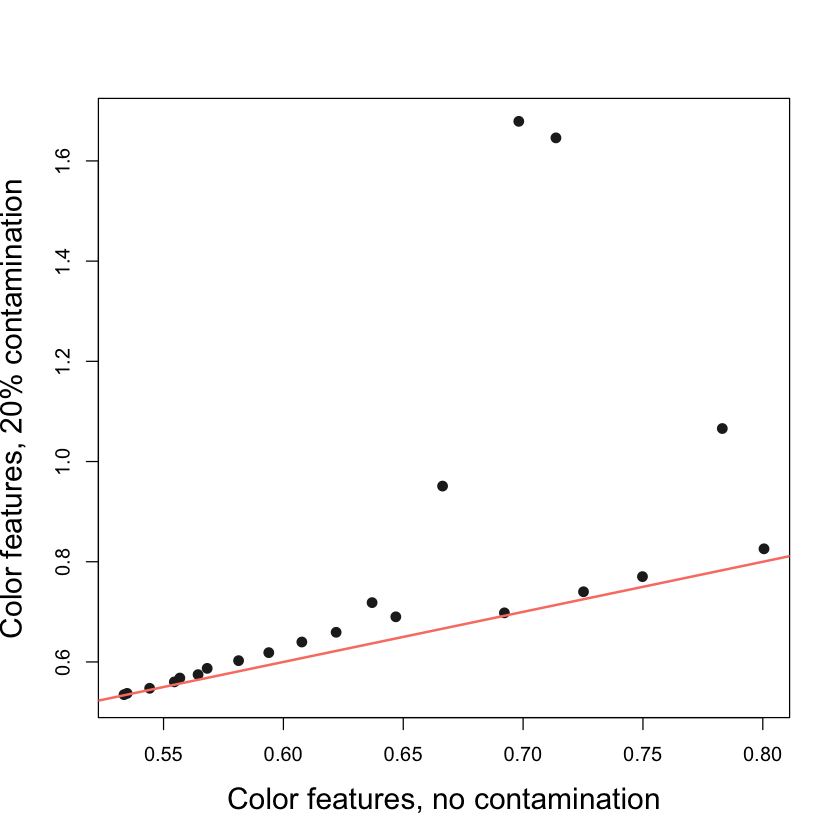}\\ 
    \end{tabular}
    \caption{Effect on our color features of adding artificial contamination to the CMD of NGC~104, from the top left to the bottom right $1\%$, $5\%$, $10\%$, and $20\%$ respectively.}
    \label{contamination_color}
\end{figure*}

\begin{figure*}
    \centering
    \begin{tabular}{cc}
    \includegraphics[width=0.49\textwidth]{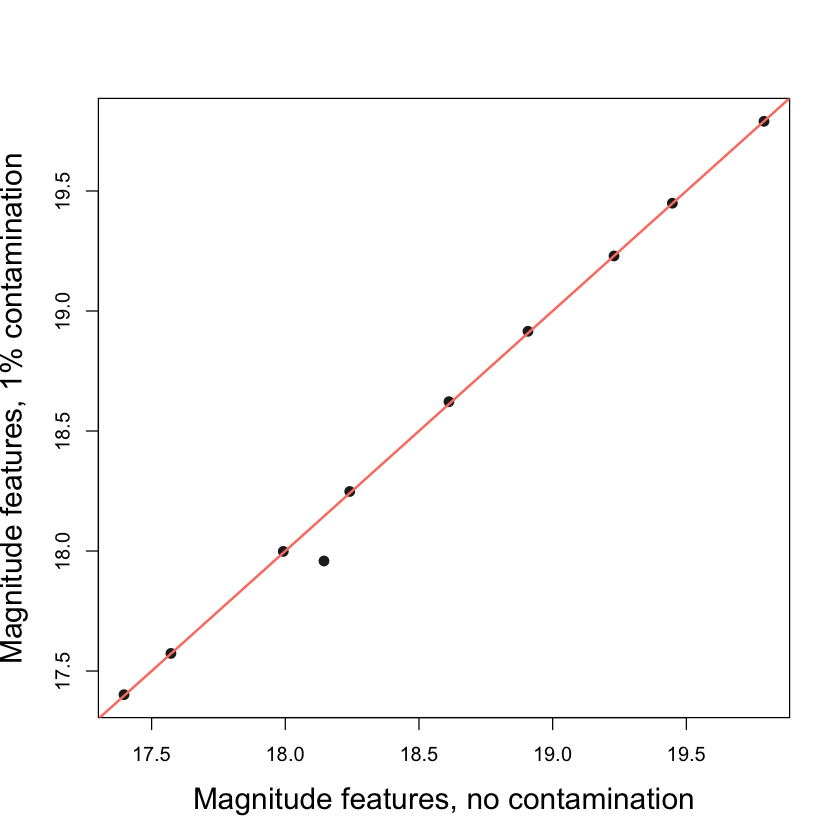} &
    \includegraphics[width=0.49\textwidth]{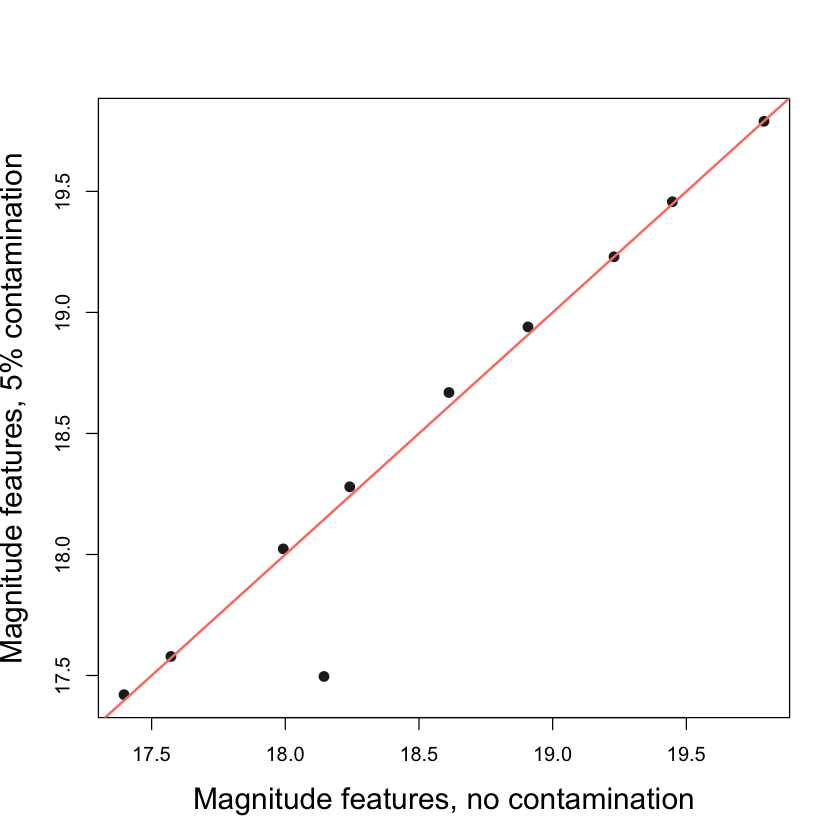} \\
    \includegraphics[width=0.49\textwidth]{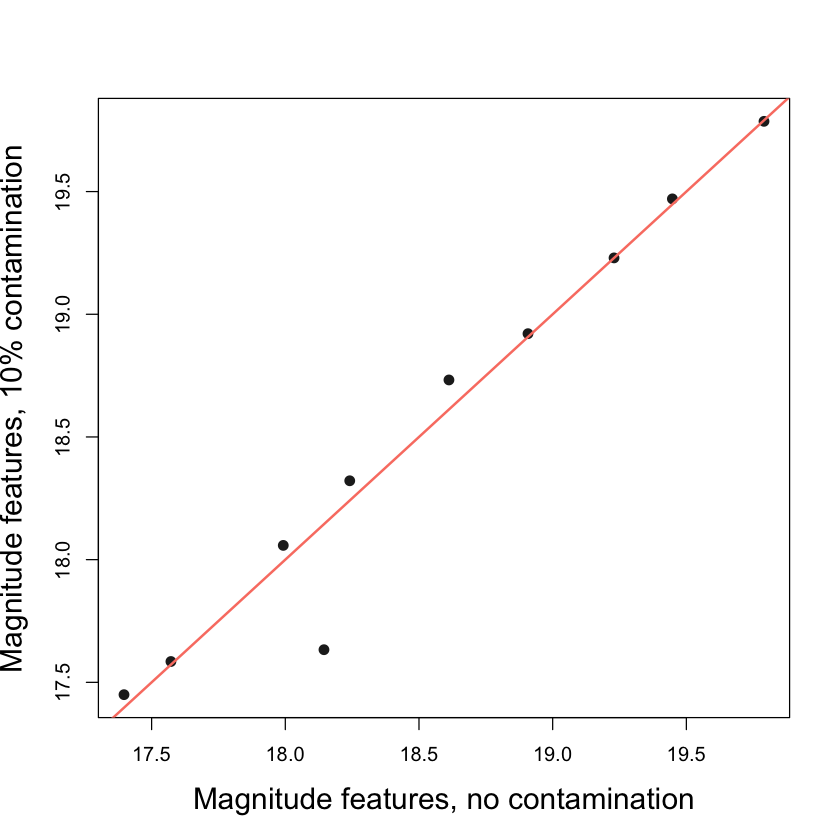}& 
    \includegraphics[width=0.49\textwidth]{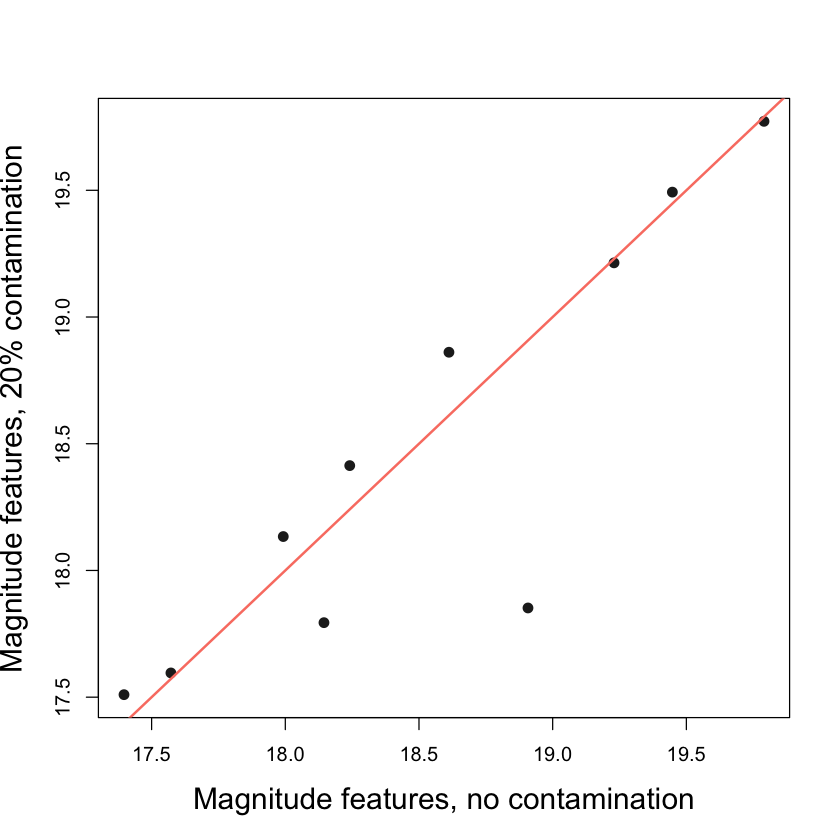}\\ 
    \end{tabular}
    \caption{Effect on our magnitude features of adding artificial contamination to the CMD of NGC~104, from the top left to the bottom right $1\%$, $5\%$, $10\%$, and $20\%$ respectively.}
    \label{contamination_mag}
\end{figure*}

\section{Results}

\begin{figure}[htbp]
\begin{center}
\centerline{\includegraphics[width=\columnwidth]{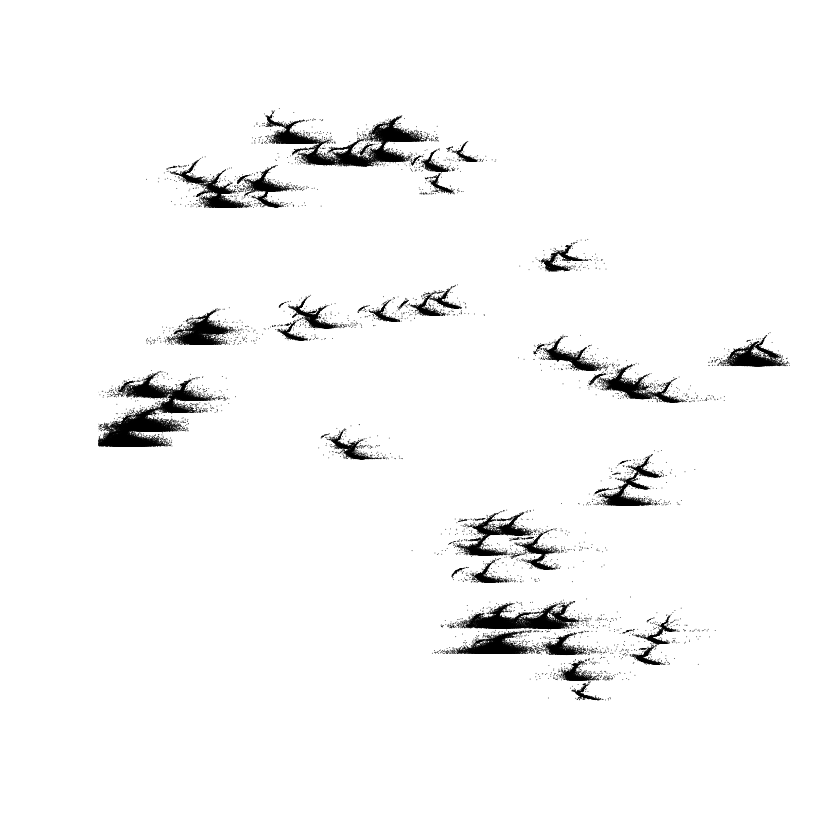}}
\caption{Outcome of t-SNE applied to quadtree feature space for our sample of $56$ CMDs. Each CMD is plotted centered on the t-SNE coordinates on the plane. Similar CMDs appear to cluster together.}
\label{tsne}
\end{center}
\end{figure}

\begin{figure*}
    \centering
    \begin{tabular}{cc}
    \includegraphics[width=0.47\textwidth]{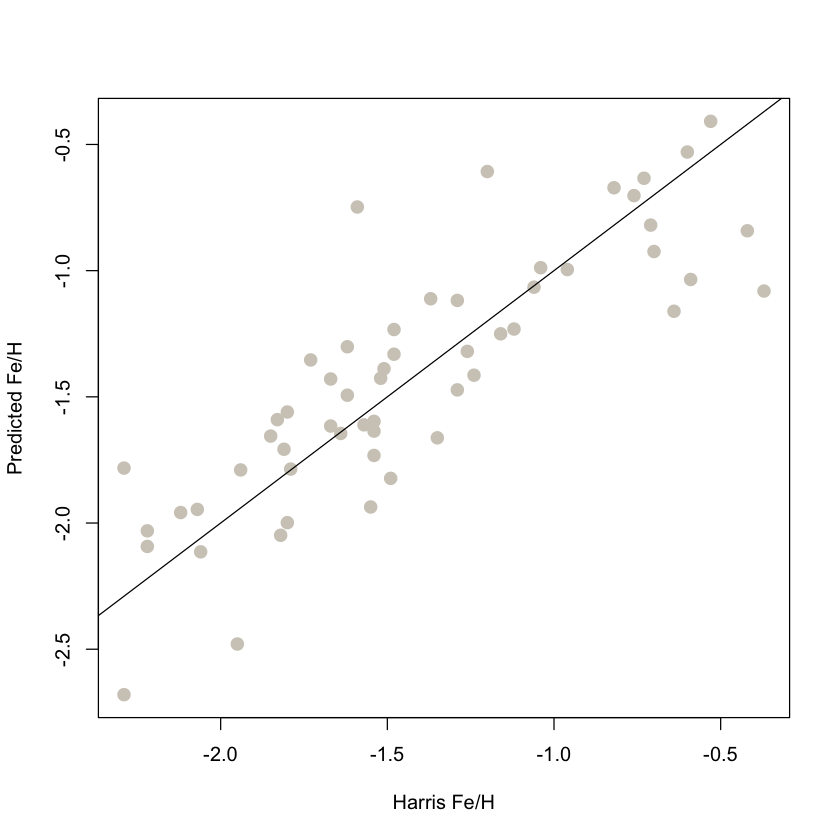} & \includegraphics[width=0.47\textwidth]{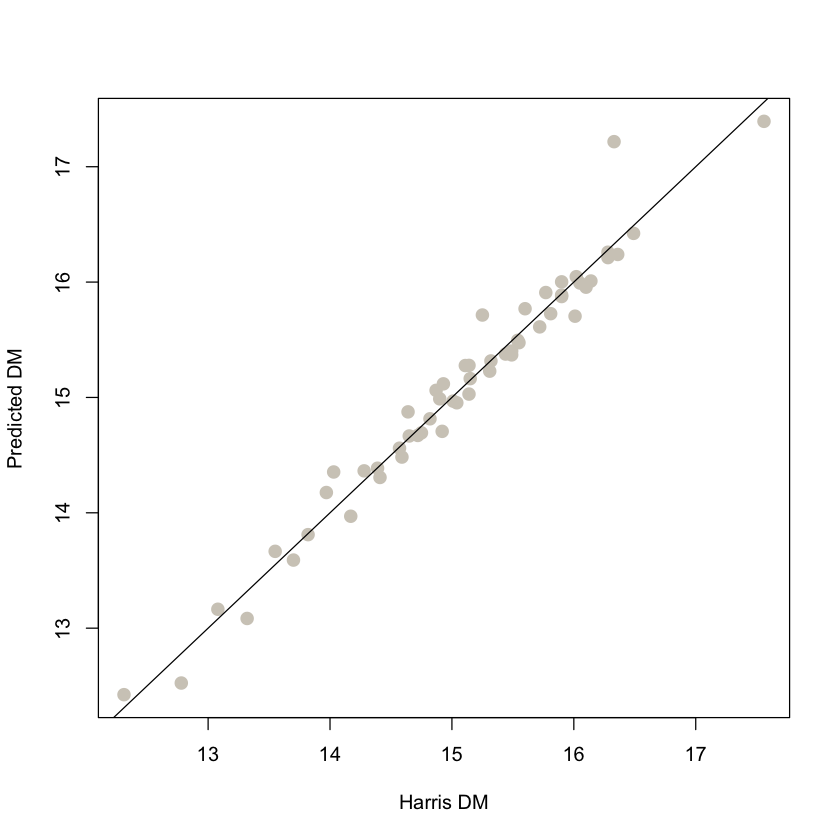} \\ 
    \end{tabular}
    \caption{Predicted versus ground truth metallicity (left panel) and DM (right panel) in leave-one-out cross-validation for a linear model trained on quadtree features of depth two, corresponding to seven features.}
    \label{supervised_FeH_DM}
\end{figure*}

\begin{itemize}
    \item Speed: We tested this empirically on a low-performance machine -an early 2015 MacBook Air. The algorithm was implemented as a recursive function in R (version 3.5.1), an interpreted language not optimized for performance. We also note in passing that the process of computing our features can be easily parallelized, because after each split we have two 2D point clouds that can be processed independently from each other; however, the code we used for the test was not parallelized, and run serially on one core only. To this end, we consider a random point cloud containing up to $10^5$ stars. The scaling appears linear in the number of stars (See Fig. \ref{scalingtime}), taking about $0.186\pm0.008$, $0.32\pm0.02$, and $0.44\pm0.02$ $\mu$s per thousand stars processed for a tree of depth $3$, $5$, and $7$ respectively. This suggests that our features could be calculated in real-time at the end of a pipeline that processes raw data into CMDs, adding little overhead.

    \item Robustness to contamination: We measure the effects of contamination by foreground stars on our features by taking the one CMD of the sample (NGC~104) and adding contaminants in varying proportions. We use \texttt{trilegal} \citep[e.g.,][]{2009A&A...498...95V} to model MW contamination in the direction of NGC~104 for this purpose. 
    In this test, we analyze the effect on the color and magnitude medians calculated after adding a fraction of contaminants corresponding to $1\%$, $5\%$, $10\%$, and $20\%$ on our features using a quadtree of depth 5 (see Fig. \ref{contamination_color} and Fig. \ref{contamination_mag}). The results show that low to moderate contamination levels disrupt only a subset of features, leaving the others intact. The disrupted features correspond to deeper levels of the quadtree, where the median has been calculated on a limited number of stars. Given that the breakpoint of the median corresponds to $50\%$ of the data, even in the worst-case scenario where an adversary can place contaminants arbitrarily, $N/2^{k+1}$ contaminants are required to affect a median at depth $k$ (having defined $k=0$ as the root of the tree containing the whole CMD) because such a median is calculated on $N/2^k$ stars. We thus expect a contamination level of at least $100/2^5 \approx 3.1$\% to be required to affect a single feature. Our test empirically demonstrates this, where contamination at $1\%$ essentially does not affect our features, and contamination at $5\%$ makes one color median and one magnitude median deviate, with the other features essentially unaffected.

    \item Robustness to photometric error: We added artificial photometric error to the CMD of NGC~104 to mimic the observational ones.

    \item Reddening and differential reddening: Since our features are the color and magnitude medians on relevant rectangular patches in the CMD plane --in the presence of a constant reddening-- they respond to this rigid translation with a constant shift. For instance, for a reddening of E(B-V)=0.1, the reddening vector in the V, B-V plane would be $(0.1, 0.31)$, so the color and magnitude features will be increased by $0.1$ and $0.31$, respectively. For differential reddening by adding random Gaussian noise to the modulus of the reddening vector while keeping its direction constant. Because our features are medians on the $N$ stars that fall in the relevant quadrant, their sampling error is related to the individual error on stellar magnitudes and colors as $\sigma_f = c \frac{\sigma_i}{\sqrt{N}}$ assuming that the individual error is normal, with $c \approx 5/4$. Since the branching of the tree at depth $k$ calculates the median color or magnitude of $M/2^k$ stars where $M$ is the total number of stars in the CMD, even at $k=6$ which is the maximum value we considered, $N \approx 16$ even if $M = 10^3$. This reduces the error by more than a factor of 3, and most points fall within the one-sigma.

    \item Dimensionality reduction of a set of GC CMDs: We performed t-SNE analysis on our features with a tree of depth $3$ on a sample of $56$ GCs. 
    Fig.~\ref{tsne} shows the t-SNE plot which displays a visually compelling structure that suggests a meaningful distribution of the data points in the feature space\footnote{Note, however, that the appearance of the plot depends on t-SNE hyperparameter \emph{perplexity}, which was set to $3$ in our case, purposefully singling out small groups of similar CMDs.}, with similar-looking CMDs often appearing near each other. 

    \item Supervised regression on a set of GC CMDs: We calculated our features with a tree of depth $3$ on a sample of $56$ GCs, resulting in $7$ features for each GC. We trained a simple linear regression model. Fig.~\ref{supervised_FeH_DM} shows the predicted [Fe/H] (left panel) and DM (left panel) on our data set, evaluating its performance in terms of rms error in leave-one-out cross-validation. Even with a simple model the scatter in predicted metallicity is $0.16$ dex and in distance modulus $0.33$ mag.

\end{itemize}

\section{Discussion and conclusions} 
\label{sec:discu}
Astronomy has recently seen a progressive increase in machine learning applications, driven by the availability of tools and the sheer volume of data. While the use of deep learning tools for the analysis of images and spectra is now commonplace, point cloud data is lagging behind. While point-cloud deep learning approaches \citep[e.g.][]{guo2020deep} can be tailored to astronomical data, astronomers doubtlessly find value in extracting tabular features from point clouds, as in the case of color-magnitude diagrams. For this application, no out-of-the-box solution is available at present.
In this work, we presented a featurization approach that is fast to compute, effective in concisely describing the data, and robust to photometric error and outliers. We demonstrated these properties empirically on actual astronomical data and articulated their theoretical underpinning where applicable. This technique overcomes the problems present in the methods used in astronomy for this kind of analysis such as the time-consuming and data handling.

\section*{Acknowledgements}
J.S-U and his work were supported by the National Agency for Research and Development (ANID)/Programa de Becas de Doctorado en el extranjero/DOCTORADO BECASCHILE/2019-72200126. 
MP acknowledges financial support from the European Union’s Horizon 2020 research and innovation programme under the Marie Sklodowska-Curie grant agreement No. 896248.
This work was partially funded by the PRIN INAF 2019 grant ObFu 1.05.01.85.14 (\emph{'Building up the halo: chemo-dynamical tagging in the age of large surveys'}, PI. S. Lucatello) 




\bibliography{example_paper}
\bibliographystyle{icml2023}


\section*{Contribution of Latin$X$ authors}
The first author is Latin$X$ and wrote part of this extended abstract and contributed to the early discussion and brainstorming that resulted in the adoption of quadtree features for CMDs. The Latin$X$ author will be the presenting author in the conference. 
\end{document}